\documentclass{llncs}

\usepackage{graphicx}
\usepackage{listingsVDM}
\usepackage{overturelanguagedef}

\begin{document}

\title{Further Progress Towards Operation Proof Obligation Generation for VDM}
\titlerunning{Further Progress Towards Operation POG for VDM}

\author{Nick Battle\inst{1} \orcidID{0009-0001-1523-4964}, Carlo Rende\inst{1} \orcidID{0009-0003-7379-6767} and Peter Gorm Larsen\inst{1} \orcidID{0000-0002-4589-1500}}
\institute{Department of Electrical and Computer Engineering, Aarhus University, Denmark, \email{nick.battle@gmail.com, crende@ece.au.dk, pgl@ece.au.dk}}

\maketitle

\begin{abstract}
The VDM formalism can ensure that models are internally consistent. Potential inconsistencies are highlighted by assertions called proof obligations. This paper is an update to the work described in \cite{Battle2025}, regarding proof obligation generation for VDM operations. We present the latest work which improves the approach to loop invariants, adds loop \emph{variants}, improves proof obligations for operation calls, including implicitly declared operations and specification statements, and adds recursive measures for operations. The new features are illustrated with examples.
\end{abstract}

\section{Introduction}
\label{sec:intro}

Formal specification methods, such as the Vienna Development Method (VDM, \cite{ISOVDM96a}), make it possible to identify all places in a model where a run-time error could occur \cite{Fitzgerald&98,Fitzgerald&05}. Proving the absence of such errors depends on the discharge of \emph{proof obligations} (POs), which are created from a model using proof obligation generation (POG) tools \cite{Aichernig&97}.

As described in \cite{Battle2025}, tool support for POG from VDM \emph{functions} is mature. However, support for the generation of POs from VDM \emph{operations} has historically been limited. This is mostly because the depth of analysis required is greater, since operations have a more complex control flow, possibly involving loops, onward operation calls, exceptions and state changes. In this paper, we describe the latest progress towards generating obligations for VDM operations.

As in \cite{Battle2025}, the new POG functionality is enabled using VDM annotations -- comments that support the POG. A new \texttt{@LoopMeasure} annotation is introduced for loop variants and recursive operation measures are supported by a new \texttt{@OperationMeasure} annotation.

In the remainder of the paper, Section \ref{sec:previous} summarizes the progress achieved in \cite{Battle2025}. Section \ref{sec:loops} describes the new work for loop obligations, then Section \ref{sec:ops} covers improvements to obligations for operation calls. Section \ref{sec:quickcheck} looks at the result of using the QuickCheck tool with the new obligations \cite{Battle2024}. Afterwards, Section \ref{sec:adequate} considers the design of adequate model constraints and Section \ref{sec:vscode} looks at GUI enhancements. Finally, Section \ref{sec:future} considers the future work.

\section{Previous Status}
\label{sec:previous}
The following principles are central to the approach described in \cite{Battle2025}:

\begin{itemize}
    \item The complexity of operation control flows is managed by analysing possible paths (ignoring loops) and then creating obligations for each possible path from the start to the point of an obligation (such as a potential division by zero).
    \item Operations can update the values of variables along a path, but obligations are pure expressions which do not have state. State changes are represented in obligations by creating nested local scopes for updated variables, hiding the value of the same name in outer scopes.
    \item Operations are defined in a module context, which can include a state vector. This is represented by an extra argument in the outermost scope of obligations, effectively quantifying over all possible states. This is more complicated for VDM++ and VDM-RT, see Section~\ref{sec:future}.
\end{itemize}

\noindent
These principles allowed the generation of many operation obligations, but the work was limited in the following regards:

\begin{itemize}
    \item The proposed approach to loop invariants did not scale well, in particular for nested loops, and there was no support for loop variants (to potentially prove termination).
    \item Operation calls from within an operation marked updated state variables as \emph{ambiguous}, meaning that their values are unknown and obligations that depend on ambiguous values could not be checked.
    \item There was no support for recursive operation measures.
    \item There was no support for exception handling, or VDM++/VDM-RT states.
\end{itemize}

\noindent
In this work, we present improvements to the first three limitations above.

\section{Loop Obligations}
\label{sec:loops}

\subsection{Invariants}
\label{sec:loopinvs}

Loop invariants have been used for formal analysis and verification for many years \cite{10.1145/363235.363259}. An invariant is a Boolean expression involving the variables that are visible to the loop, whose value is true before the loop, at the start and end of each loop iteration, and after the loop.

As introduced in \cite{Battle2025}, VDMJ provides a \texttt{@LoopInvariant} annotation. The POG uses the annotation to create POs at points before, during and after the loop annotated.

Consider the following simple loop:

\begin{vdmtt}
state Sigma of
	sv : nat
end

operations
	op: nat ==> nat
	op(a) ==
	(
		dcl local : nat := 0;
		sv := a;

		-- @LoopInvariant(sv + local = a)
		while sv > 0 do
		(
			sv := sv - 1;
			local := local + 1
		);

		return a - 1
	);
\end{vdmtt}

\noindent
This specification produces six POs:

\begin{vdmtt}
Proof Obligation 1: (Unproved)
-- check invariant before while condition
(forall a:nat, mk_Sigma(sv):Sigma &
  (let local : nat = 0 in
    (let sv : nat = a in
      ((sv + local) = a))))
\end{vdmtt}

\noindent
PO \#1 checks the invariant before the loop starts. The context at this point has the parameter bindings for \texttt{a} and \texttt{sv}, followed by the definition of \texttt{local} and the assignment to \texttt{sv}. The obligation then adds a check of the invariant in that context on the last line.
\vspace{10pt}

\begin{vdmtt}
Proof Obligation 2: (Unproved)
-- check invariant before first while body
(forall a:nat, mk_Sigma(sv):Sigma &
  (let local : nat = 0 in
    (let sv : nat = a in
      ((sv > 0) =>
        ((sv + local) = a)))))
\end{vdmtt}

\noindent
PO \#2 is very similar to \#1, but with the additional  \texttt{(sv > 0) => ...} condition that the while loop is actually entered. This represents the check that the invariant holds at the start of the first loop. (PO \#6 covers the case where the loop is not entered at all.)
\vspace{10pt}

\begin{vdmtt}
Proof Obligation 3: (Unproved)
(forall a:nat, mk_Sigma(sv):Sigma &
  (let local : nat = 0 in
    (let sv : nat = a in
      ((sv > 0) =>
        (forall local:nat, sv:nat &
          (((sv + local) = a) and (sv > 0) =>
            (sv - 1) >= 0))))))
\end{vdmtt}

\noindent
PO \#3 is a check within the body of the loop, since the \texttt{sv} value is a natural number and cannot be decremented below zero. Notice that the context now includes a \texttt{forall}, quantifying over the \texttt{local} and \texttt{sv} values, immediately qualified by the invariant and the loop condition. These are the two values changed by the loop, so the PO is saying ``for all possible changes made by the loop, if the invariant holds and the loop has not terminated, then the obligation holds''.
\vspace{10pt}

\begin{vdmtt}
Proof Obligation 4: (Unproved)
-- check invariant preserved by while body
(forall a:nat, mk_Sigma(sv):Sigma &
  (let local : nat = 0 in
    (let sv : nat = a in
      ((sv > 0) =>
        (forall local:nat, sv:nat &
          (((sv + local) = a) and (sv > 0) =>
            (let sv : nat = (sv - 1) in
              (let local : nat = (local + 1) in
                ((sv + local) = a)))))))))
\end{vdmtt}

\noindent
PO \#4 follows a similar pattern to \#3, which checks that if the loop is entered and the invariant holds at the start, then after performing the updates made by the body, the invariant still holds at the end. Together with PO \#2, this forms the inductive basis of the loop invariant.
\vspace{10pt}

\begin{vdmtt}
Proof Obligation 5: (Unproved)
(forall a:nat, mk_Sigma(sv):Sigma &
  (let local : nat = 0 in
    (let sv : nat = a in
      ((sv > 0) =>
        (forall local:nat, sv:nat &
          (((sv + local) = a) and (not (sv > 0)) =>
            (a - 1) >= 0))))))
\end{vdmtt}

\noindent
PO \#5 is caused by the \texttt{return a - 1} statement after the loop. This illustrates how the POG adds context for obligations immediately after the loop, effectively saying that we know the invariant still holds and the loop condition is no longer met (i.e. the loop has terminated).
\vspace{10pt}

\begin{vdmtt}
Proof Obligation 6: (Unproved)
(forall a:nat, mk_Sigma(sv):Sigma &
  (let local : nat = 0 in
    (let sv : nat = a in
      ((not (sv > 0)) =>
        (-- Did not enter loop at 13:9
          (((sv + local) = a) =>
            (a - 1) >= 0))))))
\end{vdmtt}

\noindent
The final PO \#6 covers the path where the loop was not entered. In this case, the obligation still holds after the loop, even though no changes were made.

\subsection{Other Loop Types}
\label{sec:otherloops}

\emph{While-loops} are the simplest case, but VDM also has three \emph{for-loop} types. The POs produced follow the same overall principle as \emph{while-loops}, with the following differences:

\begin{itemize}
    \item \emph{for-index} loops are of the form \texttt{for x = A to B by S do}. In this case, the PO before the loop is the case when \texttt{x = A}, because if the loop was an equivalent \emph{while-loop}, the \texttt{x} variable would be set to \texttt{A} outside, followed by a \texttt{while x <= B do} loop. In our example, the loop quantifier for \texttt{for x = 1 to a do}, would quantify over the \texttt{x} values, as well as the loop state changes, as follows\footnote{A \texttt{by S} clause would add \texttt{((x - A) \textbf{rem} S) = 0} to the constraint, and swap A/B comparisons if negative.}:
\begin{vdmtt}
((1 <= a) =>
  (forall x:nat1, local:nat, sv:nat &
    (((x >= 1) and (x <= a)) and ((sv + local) = a) =>
\end{vdmtt}
    \vspace{10pt}

    \item \emph{for-set} loops are of the form \texttt{for all x in set S do}. The difficulty here is that sets are not ordered, therefore the order in which the elements are processed is not defined. To handle this, the PO uses a \emph{ghost variable} which holds the elements processed so far; the invariant may reason about the ghost variable. By default, the ghost is called \texttt{DONE\_n\$}, where n is the line number of the loop\footnote{The ghost can be given a more sensible name, using a second argument to the annotation.}. The quantifier in for \texttt{for all x in \{1, ..., a\} do} would be as follows:
\begin{vdmtt}
(({1, ..., a} <> {}) =>
 (forall x:nat1, local:nat, sv:nat, DONE_13$:set of nat1 &
    (DONE_13$ psubset {1, ..., a})
    and (x in set ({1, ..., a} \ DONE_13$))
    and ((sv + local) = a) =>        
\end{vdmtt}
    Then, for the obligation that checks the invariant at the end of the loop, the ghost variable is first updated to include the current \texttt{x} value before checking the invariant:
\begin{vdmtt}
(let DONE_13$ : set of nat1 = DONE_13$ union {x} in
  ((sv + local) = a))
\end{vdmtt}
    \vspace{10pt}

    \item \emph{for-seq} loops are of the form \texttt{for p in S do}, where \texttt{S} is a sequence. Although sequences are ordered, the approach taken here is similar to \emph{for-set} loops, using a ghost sequence to hold the elements done so far, allowing the invariant to reason about them. The quantifier in \texttt{for x in list do}, and the update of the ghost, would be as follows (using the length of the ghost in the invariant, instead of \texttt{local}):
\begin{vdmtt}
((list <> []) =>
 (forall x:nat1, DONE_13$:seq of nat1, local:nat, sv:nat &
   (DONE_13$ ^ [x] = list(1, ..., len DONE_13$ + 1))
   and ((sv + (len DONE_13$)) = a) =>
   ...
   (let DONE_13$ : seq of nat1 = DONE_13$ ^ [x] in
     ((sv + (len DONE_13$)) = a)
\end{vdmtt}
\end{itemize}

\subsection{Variants}
\label{sec:loopvars}

Loop variants can be used in \emph{while-loops} to verify that loops will terminate. This is not necessary with \emph{for-loops}, which always terminate. Variants are introduced via the \texttt{@LoopMeasure}\footnote{Because of the similarity to function measures and to the \texttt{@OperationMeasure} annotation. The name \texttt{@LoopVariant} might be confused with \texttt{@LoopInvariant}.} annotation.

A loop variant is an expression whose value is a natural number, which decreases towards zero for every iteration of the loop. In the example above, the \texttt{sv} variable would act as a valid measure, so the loop would be annotated like this:
\begin{vdmtt}
    -- @LoopInvariant(sv + local = a)
    -- @LoopMeasure(sv)
    while sv > 0 do
    ...
\end{vdmtt}
The variant produces an additional obligation, saying that the measure (held in a ghost variable called \texttt{LOOP\_n\$}) at the start of a loop iteration is greater than the measure at the end:
\begin{vdmtt}
Proof Obligation 1: (Unproved)
-- check measure after each while body
(forall a:nat, mk_Sigma(sv):Sigma &
  (let local : nat = 0 in
    (let sv : nat = a in
      (forall local:nat, sv:nat &
        (((sv + local) = a) and (sv > 0) =>
          (let LOOP_13$ : nat = sv in
            (let sv : nat = (sv - 1) in
              (let local : nat = (local + 1) in
                sv < LOOP_13$))))))))
\end{vdmtt}

\subsection{Runtime Checks}
\label{sec:runtimeloops}
Loop invariants and loop measures are both checked by the VDMJ interpreter at runtime, so even if POs are not formally discharged, it may be possible to find invariant/measure violations via traditional test methods. Ghost variables appear in the runtime context as variables.

\section{Operation Calls}
\label{sec:ops}

\subsection{Simple Calls}
\label{sec:opcalls}
Simple operation calls do not return a value. These cause a problem in POs because a called operation can make arbitrary updates to state data, which must be represented in the PO context. But unlike (say) a direct assignment to state, we cannot easily calculate the state updates that an operation will make.

The approach taken in \cite{Battle2025} was to naively calculate the variable update set, and mark these as \emph{ambiguous}, meaning that their values are unknown after the call. If the PO subsequently reasoned about these values, it was marked as ``UNCHECKED''.

The new approach calculates a more accurate transitive closure of the variables that can be updated by an operation call. The possible values of these variables are then quantified over, before being qualified by any pre/post constraints on the operation being called. For example:

\begin{vdmtt}
state Sigma of
    sv : nat
end

operations
    op: nat ==> nat
    op(x) == (
        op2(x + 1);
        return 1/sv   -- Proof obligation here! (sv <> 0)
    );

    op2(a:nat)        -- NOTE: no return value
    ext wr sv         -- Changes "sv"
    pre a > sv
    post sv > a;


Proof Obligation 1: (Unproved)
(forall x:nat, mk_Sigma(sv):Sigma &
  pre_op2((x + 1), mk_Sigma(sv)))

Proof Obligation 2: (Unproved)
(forall x:nat, mk_Sigma(sv):Sigma &
  (let $oldState = mk_Sigma(sv) in
    (forall sv:nat &    -- Because op2 has "ext wr sv"
      pre_op2((x + 1), $oldState)
      and post_op2((x + 1), $oldState, mk_Sigma(sv)) =>
        sv <> 0)))      -- PO for the return 1/sv
\end{vdmtt}

\noindent
The first PO checks that the precondition for \texttt{op2} is always met. The second PO records the ``old'' state, before quantifying over possible values of the affected state, \texttt{sv}. This is then qualified by both the precondition over the old state, and the postcondition over the old state plus the new updated state. If all of that context is met, then we are reasoning about a valid scenario and so the final obligation must hold in that case.

The pattern of quantifying over the possibilities, then qualifying them with the constraints, is similar to the loop invariant checks, where we quantify over the possible state updates in the loop, and qualify them with the loop invariant and loop condition.

This approach also allows POs to be generated for implicit operation definitions (as the example above) or for specification statements, which are similar but inline.

Note that an operation call may be to another module, which has its own opaque state. That may then call back to the original module and update its state\footnote{Note that we do not know which operation does this, so we cannot use its arguments or postcondition to qualify the possible update values.}. In general, we are only concerned about the possible state updates in the local module, but we do have to allow a remote module to have an arbitrary state too. For example, if we call an operation in module \texttt{B}, with opaque state \texttt{B`Gamma}, from local module \texttt{A} with state \texttt{Sigma}, the POs look like this\footnote{POs have global visibility of all modules' state names, though remote state structures remain opaque.}:

\begin{vdmtt}
Proof Obligation 1: (Unproved)
(forall x:nat, mk_Sigma(sv):Sigma &
  (forall $oldState:B`Gamma &
    B`pre_op2((x + 1), $oldState)))

Proof Obligation 2: (Unproved)
(forall x:nat, mk_Sigma(sv):Sigma &
  (forall $oldState:B`Gamma, $newState:B`Gamma &
    (-- Change local state here too? eg. forall sv:nat &
      B`pre_op2((x + 1), $oldState)
      and B`post_op2((x + 1), $oldState, $newState) =>
        sv <> 0)))
\end{vdmtt}

\noindent
Note that the PO knows nothing about \texttt{Gamma} values and does not try to track them. However, it quantifies over the possible \texttt{Gamma} states when calling operations in the \texttt{B} module. Therefore, the onus is on the specifier to write strong pre/post conditions for the operation, so that the valid \texttt{Gamma} states are identifiable. This is discussed further in Section \ref{sec:adequate}.

\subsection{Return Values}
\label{sec:returns}
As well as updating state, operation calls can return values which are used in calculations by the caller. The ISO VDM-SL Standard \cite{ISOVDM96a} limits the places where operations can be called to yield a value, since the order of operation evaluation is significant, and the order of calls in an arbitrary expression is not defined. However, most VDM tools relax this constraint in anticipation of VDM++/RT, which use operation calls with less rigour.

The problem for the POG is firstly to find and extract the operation calls from an expression. It must then work out the order of evaluation of those operation calls, before adding the possible results to the values quantified over as a result of making each call. The possible results must then be substituted back into the original expression to continue with the context path.

For example, if a specification makes two calls \emph{to the same operation} and adds the results together:
\begin{vdmtt}
def rv = op2(x + 1) + op2(x * 2) in return 1/rv
\end{vdmtt}
Then the important obligation looks like this:

\begin{vdmtt}
Proof Obligation 1: (Unproved)
(forall x:nat, mk_Sigma(sv):Sigma &
  (let $oldState = mk_Sigma(sv) in
    (forall sv:nat, $op2:nat &       -- Possible sv + result
       pre_op2((x + 1), $oldState)
       and post_op2(
         (x + 1), $op2, $oldState, mk_Sigma(sv)) =>
      (let $oldState = mk_Sigma(sv) in
        (forall sv:nat, $op2$1:nat & -- Possible sv + result
           pre_op2((x * 2), $oldState)
           and post_op2(
             (x * 2), $op2$1, $oldState, mk_Sigma(sv)) =>
          (let rv = ($op2 + $op2$1) in  -- Substituted back
            rv <> 0))))))
\end{vdmtt}

\noindent
The \texttt{\$op2} and \texttt{\$op2\$1} variables are the return values from the two calls to \texttt{op2}. This naming pattern extends to an arbitrary number of calls to the same operation in an expression. Return values are quantified over in the PO context, along with the possible updates to \texttt{sv}, qualified by the operation's precondition and postcondition.

Note that the extracted calls to \texttt{op2} are performed in the order shown. This is because VDMJ performs the addition in left-to-right order. In general, extracted operation calls have to appear in tool evaluation order in the PO context. Finally, for every valid combination of results and state updates, the results are added together to set \texttt{rv} before the final obligation check.

This approach does permit most operation calls with return values to be handled by the POG, but some cases are not possible, such as when an expression generates an unknown number of calls, like \texttt{\{ op2(a) | a in set S \}} -- here we cannot statically determine how many calls are made by the set comprehension, and the resulting POs are marked as ``UNCHECKED''.

\subsection{Recursive Measures}
\label{sec:measures}
Operations may be recursive in VDM, similar to functions, though operations do not have to terminate\footnote{For example, a \texttt{while true do skip} loop does not terminate.}. Recursive functions can define a \texttt{measure} to verify that recursion will terminate. We enable a similar feature for operations via an \texttt{@OperationMeasure} annotation. For example\footnote{The subtle error in this operation is picked up in Section \ref{sec:quickcheck}.}:

\begin{vdmtt}
state Sigma of
    sv : nat
end

operations
    -- @OperationMeasure(a)
    fac: nat ==> nat
    fac(a) ==
        if a = 1
        then return 1
        else return a * fac(a-1);

Proof Obligation 1: (Unproved)
(forall a:nat, mk_Sigma(sv):Sigma &
  (let MEASURE_6$ : nat = a in
    (not (a = 1) =>
      MEASURE_6$ > measure_fac(a - 1, mk_Sigma(sv)))))
\end{vdmtt}

\noindent
The presence of the \texttt{@OperationMeasure} annotation causes the creation of a \texttt{measure\_fac} function, which has the same signature as \texttt{pre\_fac}, but which evaluates the measure expression provided.

The obligations generated are very similar to those for recursive function calls, but with the addition of the state argument. The temporary variable holding the measure includes the line number of the operation.

This direct recursive example is a special case of a more general pattern where mutually recursive cycles of operation calls are required to define a measure which decreases at each step of the cycle. For example, if the factorial example is divided into two operations, in two modules, which call each other alternately, the obligations would be as follows:

\begin{vdmtt}
Proof Obligation 1: (Unproved)
(forall a:nat, mk_Sigma(sv):Sigma &
  (let MEASURE_11$ : nat = a in
    (not (a = 1) =>
      (forall $oldState:B`Gamma &
        MEASURE_11$ > B`measure_fac(a - 1, $oldState)))))

Proof Obligation 2: (Unproved)
(forall b:nat, mk_Gamma(xv):Gamma &
  (let MEASURE_28$ : nat = b in
    (not (b = 1) =>
      (forall $oldState:A`Sigma &
        MEASURE_28$ > A`measure_fac(b - 1, $oldState)))))
\end{vdmtt}

\noindent
Note that the state of the remote modules is opaque to each caller, so a quantifier is added to allow the state to potentially affect the measure. In the direct recursive case, the PO can construct the current \texttt{Sigma} state directly.

\subsection{Runtime Checks}
\label{sec:runtimeops}
As with loop invariants, operation measures are checked by the VDMJ interpreter at runtime, so even if POs are not formally discharged, it may be possible to find measure violations via traditional test methods.

\section{QuickCheck}
\label{sec:quickcheck}
The new obligations produced by this work are checkable by the QuickCheck tool \cite{Battle2024}, which complements VDMJ's POG. For example, the recursive factorial operation in Section \ref{sec:measures} produces the following results:

\begin{vdmtt}
> qc
PO #1, FAILED in 0.009s
Counterexample:
fac: recursive operation obligation at line 11:25
(forall a:nat, mk_Sigma(sv):Sigma &
 --> sv = 0, a = 0
  (let MEASURE_7$ : nat = a in
   --> MEASURE_7$ = 0
    (not (a = 1) =>
      MEASURE_7$ > measure_fac(a - 1, mk_Sigma(sv)))))
       --> Error 4065: Value -1 is not a nat at line 4:20

PO #2, FAILED in 0.001s
Counterexample:
fac: subtype obligation at line 11:30
(forall a:nat, mk_Sigma(sv):Sigma &
 --> sv = 0, a = 0
  (let MEASURE_7$ : nat = a in
    (not (a = 1) =>
      (a - 1) >= 0)))
      --> returns false
> 
\end{vdmtt}

\noindent
In this particular example, the operation fails if the argument passed is zero, because the recursive exit check is \texttt{a = 1} rather than \texttt{a <= 1}. This can be corrected by changing the test or adding a precondition.

The obligations created by this work reduce the number of ``UNCHECKED'' results that the POG produces, allowing QuickCheck to analyse more of them. As the work has progressed, the POG has been repeatedly tested on a large corpus of VDM-SL specifications that are distributed with the tool, producing over 7000 obligations. In \cite{Battle2025}, the POG marked 9.6\% of these as ``UNCHECKED''; with the latest work, that figure is down to 2.4\%.

\section{Adequate Constraints}
\label{sec:adequate}

The new obligations, for both loops and operation calls, depend on invariants and postconditions to qualify the large number of possibilities, representing all of the state changes that could occur. The ability to discharge obligations which follow loops or operation calls therefore depends on the \emph{adequacy} of those constraints: they must provide enough information to enable a theorem prover to discharge the obligations.

Judging how to balance the strength of constraints can be a challenge, even for experienced users. However, the need to provide enough information to fulfil obligations is a useful guide. It is our belief that tool support aimed at helping to design constraints, in the context of the obligations they affect, would be very productive.

To that end, a new code lens has been produced for the VDM-VSCode tool\footnote{See https://github.com/overturetool/vdm-vscode.}, which adds a clickable widget above \texttt{pre}, \texttt{post} and \texttt{inv} constraints. Clicking the widget causes the POG view to display the obligations which are \emph{directly affected} by that constraint. The problem for the specifier is to design a constraint such that it meets the specification's requirement, but provides enough information to discharge the obligations that it affects.

The current approach to operation POG produces much more complicated obligations than those generated for functions. The discharge of complicated obligations can be a challenge. To manage this complexity, an approach based on variable slicing should be considered, building on \cite{oda2024}.

\section{VDM-VSCode Support}
\label{sec:vscode}

The improvements to operation POG described in this paper are supported by corresponding enhancements to the VDM-VSCode tool.

The goal of these changes is to tighten the integration between the POs generated, the QuickCheck analysis described in Section \ref{sec:quickcheck}, and the model constraints discussed in Section \ref{sec:adequate}.

\subsection{Dependent Proof Obligations}
As discussed in Section \ref{sec:adequate}, the discharge of POs depends critically on the adequacy of invariants, preconditions and postconditions. To support this process, a new \emph{CodeLens}\footnote{A CodeLens is a clickable item appearing in the source editor that allows some action to be performed.} is introduced in the editor, appearing over the constraints.

Selecting this CodeLens filters the PO view to show only those obligations directly affected by the chosen constraint. This helps the specifier to reason locally about the impact of a constraint and to refine it until the dependent obligations can be discharged.

\begin{figure}
    \centering
    \includegraphics[width=\linewidth]{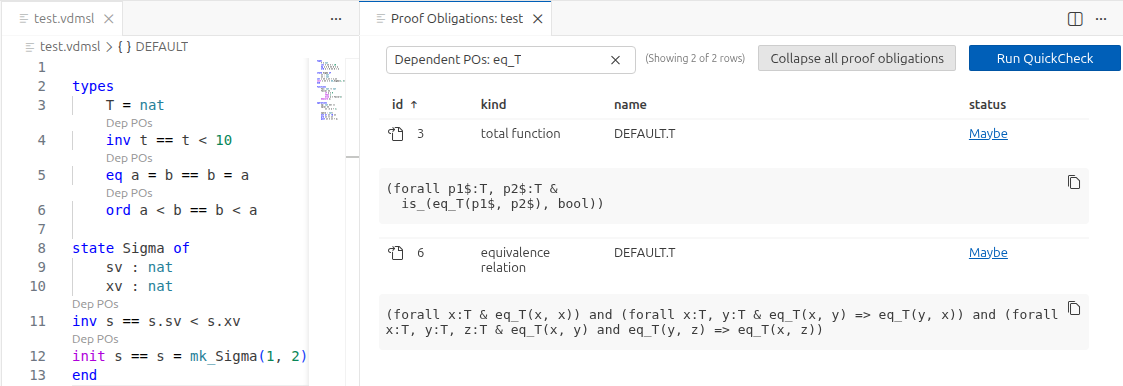}
    \caption{CodeLens-based filtering of dependent proof obligations}
    \label{fig:dependent-pos}
\end{figure}

As shown in Figure \ref{fig:dependent-pos}, selecting the CodeLens filters the PO panel to display only the obligations directly influenced by the chosen constraint.

\subsection{QuickCheck Result Presentation}
The VDM-VSCode user interface has been updated to improve the presentation of QuickCheck results within the PO panel. Rather than displaying the details provided by QuickCheck in a dedicated sub-panel, these are now provided via contextual hover elements associated with each obligation. This keeps the panel compact while still making diagnostic information immediately available.

For obligations that include a launchable counterexample or witness evaluation, clicking the obligation status navigates directly to the corresponding source location in the editor, where a CodeLens allows the counterexample/witness to be launched. This strengthens the connection between failed or provable obligations and the related constructs, making debugging more efficient.

\subsection{Missing Obligations}
As described in Section \ref{sec:previous}, obligations are generated by analysing control flow paths and producing one obligation per feasible path. In specifications with complex control flows, the number of alternative paths can quickly become unmanageable.

The POG therefore limits the number of generated obligations per definition. When this limit is reached, additional obligations are not produced and the model is reported as having \emph{missing proof obligations}.

The VDM-VSCode interface highlights these cases with warnings shown above the obligation table, and a button to navigate to each definition concerned. In addition, an \emph{InlayHint}\footnote{An InlayHint appears as a warning triangle in the editor, which expands to a detailed explanation on a hover.} is added in the editor to give contextual information about the problem. This helps the user to find and simplify complex definitions.

\begin{figure}
    \centering
    \includegraphics[width=\linewidth]{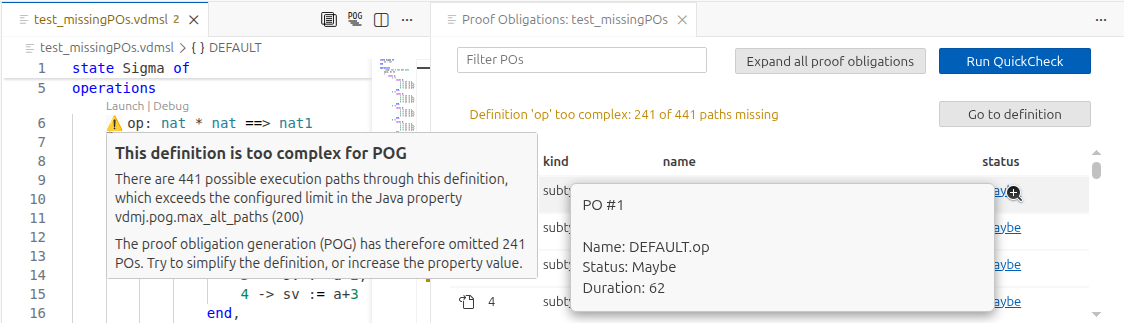}
    \caption{Presentation of QuickCheck results via contextual hover and missing proof obligations warnings}
    \label{fig:qc-missing-pos}
\end{figure}

Figure \ref{fig:qc-missing-pos} illustrates the presentation of QuickCheck diagnostics via contextual hover and missing POs warnings within the PO panel and editor.

\section{Future Work}
\label{sec:future}
Noting the limitations in Section \ref{sec:previous}, the current work has addressed loop invariants and operation calls, including measures. The remaining work items from \cite[Section 7, Future Work]{Battle2025} are:

\begin{itemize}
\item Statements that handle exceptions (\texttt{always}, \texttt{trap} and \texttt{tixe}) cause control flows that are currently too complex to handle.
\vspace{10pt}

\item Specifications that include variable hiding can easily confuse the POG and produce invalid POs.
\vspace{10pt}

\item The correct operation of the POG itself must be determined. This should ultimately be linked to a proof theory for VDM operations.
\vspace{10pt}

\item The complex state and control flows of VDM++ and VDM-RT cause a host of problems that have yet to be solved.

\end{itemize}

\noindent
In addition, as noted in Section~\ref{sec:returns}, we cannot yet deal with multiple operation calls in an expression where the order of those calls cannot be statically determined. Furthermore, Section \ref{sec:opcalls} noted that inter-module calls cannot yet determine how local state is updated, leading to under-qualified POs.

As mentioned in Section \ref{sec:adequate}, it is our belief that useful progress can be made by considering tool support for the creation of model constraints, given the context of the POs that need to be discharged under those constraints. And the use of variable slicing may be able to simplify POs.

\vspace{8pt}
\paragraph{\textbf{Acknowledgements}}
We are grateful for the support of the European Union, Aarhus University, Newcastle University and the Grundfos Foundation. We also thank the reviewers for their valuable feedback on the original version of this paper.

\bibliographystyle{unsrt}  
\bibliography{au.bib}

\end{document}